\documentstyle[12pt]{article}
\newlength{\abstractwidth}
\renewcommand{\thanks}[1]{\footnote{#1}} 

\newcommand{\be}{\begin{equation}}
\newcommand{\bea}{\begin{eqnarray}}
\newcommand{\eea}{\end{eqnarray}}
\newcommand{\ee}{\end{equation}}
\newcommand{\<}{\langle}
\renewcommand{\>}{\rangle}
\def\ba{\begin{eqnarray}}
\def\ea{\end{eqnarray}}

\def\F{{\cal F}}

\def\N{{\cal N}}
\def\O{{\cal O}}

\def\X{{\cal X}}

\def\Im{{\rm Im}}

\def\det{{\rm det}}
\def\sdet{{\rm sdet}}

\def\half{ {1\over 2}}

\def\p{\partial}
\def\pz{\partial _z}

\def\pw{\partial _w}

\def\tet{\vartheta}

\def\chiz{{\chi _{\bar z}{} ^+}}
\def\chiw{{\chi _{\bar w}{} ^+}}
\def\chiu{{\chi _{\bar u}{} ^+}}
\def\chiv{{\chi _{\bar v}{} ^+}}
\def\chix{{\chi _{\bar x}{} ^+}}

\begin{document}
\baselineskip=16pt

\begin{flushright}
UCLA/01/TEP/24 \\
Columbia/Math/01
\end{flushright}

\bigskip

\begin{center}
{\Large \bf TWO-LOOP SUPERSTRINGS \ I } \\
\bigskip
{\large \bf Main Formulas}
\footnote{Research supported in part by National Science Foundation
grants PHY-98-19686 and DMS-98-00783, and
by the Institute for Pure and Applied Mathematics under
NSF grant DMS-98-10282.}

\bigskip\bigskip

{\large Eric D'Hoker$^a$ and D.H. Phong$^b$} \\ 

\bigskip

$^a$ \sl Department of Physics and \\
\sl Institute for Pure and Applied Mathematics (IPAM) \\
\sl University of California, Los Angeles, CA 90095 \\
$^b$ \sl Department of Mathematics \\ 
\sl Columbia University, New York, NY 10027

\end{center}

\bigskip\bigskip

\begin{abstract}
 
An unambiguous and slice-independent formula for the two-loop 
superstring measure on moduli space for even spin structure
is constructed from first principles. The construction
uses the super-period matrix as moduli invariant under worldsheet
supersymmetry. This produces new subtle contributions to the
gauge-fixing process, which eliminate all the ambiguities plaguing
earlier gauge-fixed formulas.

The superstring measure can be computed explicitly and a simple expression in terms of modular forms is obtained. For fixed spin structure, the measure exhibits
the expected behavior under degenerations of the surface. The measure
allows for a unique modular covariant GSO projection. Under this GSO
projection, the cosmological constant, the 1-, 2- and 3- point functions
of massless supergravitons all vanish pointwise on moduli space without
the appearance of boundary terms. A certain disconnected part of the
4-point function is shown to be given by a convergent, finite integral 
on moduli space. A general slice-independent formula is given for the 
two-loop cosmological constant in compactifications with central
charge $c=15$ and $\N=1$ worldsheet supersymmetry in terms of the data of
the compactification conformal field theory. 

In this paper, a summary of the above results is presented with detailed
constructions, derivations and proofs to be provided in a series of
subsequent publications. 

\end{abstract}

\vfill\eject

\baselineskip=15pt
\setcounter{equation}{0}
\setcounter{footnote}{0}

\vfill\eject

\section{Introduction}
\setcounter{equation}{0}

Despite great advances in superstring theory,
multiloop amplitudes are still unavailable,
almost twenty years after the derivation of the
one-loop amplitudes by Green and Schwarz for Type II strings
\cite{gs82} and by Gross et al.
for heterotic strings \cite{ghmr86}. The main  obstacle is the presence
of  supermoduli for worldsheets of non-trivial topology 
\cite{fms,superm}.  Considerable efforts had been made by many authors
in order to overcome this obstacle, and a chaotic situation ensued, with
many competing prescriptions proposed in the literature.
These prescriptions drew from a variety
of fundamental principles such as
BRST invariance and the picture-changing
formalism \cite{fms,vv}, descent equations and Cech
cohomology \cite{v}, modular invariance \cite{mhns},
the light-cone gauge \cite{mand}, the global geometry of the
Teichmueller curve \cite{ams}, the unitary gauge \cite{lp},
the operator formalism \cite{opf},
group theoretic methods \cite{neveu}, factorization
\cite{fact}, and algebraic supergeometry \cite{asg}.
However, the basic problem was that gauge-fixing required a
local gauge slice, and the prescriptions
ended up depending on the
choice of such slices, violating gauge invariance. At the most pessimistic 
end, this raised the undesirable possibility that superstring
amplitudes could be ambiguous \cite{ars}, and that it may be 
necessary to  consider other options, such as the Fischler-Susskind
mechanism~\cite{ln}.

\medskip

In \cite{dp88} and \cite{dp89}, we had suggested
that the difficulties encountered in the earlier prescriptions 
could be the result of improper gauge-fixing procedures which 
did not respect worldsheet local supersymmetry. To address this
difficulty, we had outlined a new gauge-fixing procedure based on
projecting supergeometries to their super period matrices instead of
their underlying bosonic geometries. Unlike the projection
to the bosonic geometries, the projection
to the super period matrix
descends to a projection of superconformal
structures, since the super period matrix is invariant under local
worldsheet supersymmetry. It is well defined for any genus~$h$.

\medskip

In this paper,
we implement this new gauge-fixing procedure for genus $h=2$. 
This is the lowest loop order where supermoduli must be confronted
in all scattering amplitudes. We shall concentrate on the case of even
spin structures since odd spin structure contributions are absent for
the cosmological constant and scattering amplitudes with 4 or
fewer states. 

\medskip

$\bullet$ 
We obtain a gauge-fixed formula $d\mu[\delta](\Omega)$
for the contribution to the superstring measure of each even spin
structure $\delta$, which is {\sl independent of the choice of gauge
slice}.  In particular, the ambiguities plaguing the earlier 
prescriptions have now disappeared;

$\bullet$ For each $\delta$, $d\mu[\delta](\Omega)$ transforms
covariantly under modular transformations.
There is a unique assignment of relative phases $\eta _{\delta}$
so that $\sum_{\delta}\eta _{\delta}d\mu[\delta](\Omega)$
is a modular form, and hence a unique way of
implementing the Gliozzi-Scherk-Olive (GSO) projection;

$\bullet$ The superstring measure, when summed over all $\delta$,
vanishes {\sl point by point} on moduli space, and not just up to a total
derivative, as in earlier prescriptions. In particular, the cosmological
constant vanishes. This is the 2-loop generalization of the
Jacobi identity. Remarkably, it is not a consequence of genus
2 Riemann identities. Instead, it is equivalent to the identity, special
to genus 2, that any modular form of weight 8 must be proportional to the 
square of the unique modular form of weight 4.

$\bullet$ Similarly, the 1,2,3 point functions for the scattering
of the supergraviton multiplet vanish by a variety of novel identities. 

$\bullet$ The 4-point function may be evaluated explicitly in terms of
modular forms. For a certain disconnected part of the 4-point function,
explicit formulas are presented here; they are manifestly
finite, in the regime of purely imaginary Mandelstam variables. (As
is well known \cite{dp94}, the other regimes are  only accessible by
analytic continuation.)

$\bullet$ Finally, we provide a simple slice independent formula for the
even spin structure superstring measure and cosmological constant for
general compactifications with matter central charge $c=15$ and
$\N=1$ worldsheet supersymmetry.

\medskip

Since the derivation of all of these results is quite lengthy,
we provide here the main formulas, leaving the detailed
treatment to a forthcoming series of papers.

\section{The Supermoduli Space Measure}
\setcounter{equation}{0}

In the Ramond-Neveu-Schwarz (RNS) formulation \cite{rns}, the worldsheet
for superstring propagation in $10$-dimensional Minkowski space-time at
loop order $h$ is a compact surface $\Sigma$ of genus $h$, equipped with a
supergeometry
$E_M{}^A, \Omega_M$ obeying the Wess-Zumino torsion constraints
\cite{dp88,superg}. The non-chiral  vacuum-to-vacuum functional
integral for fixed even spin structure $\delta$ (before the GSO
projection) is
\be
{\bf A}[\delta]
=
\int DE_M{}^AD\Omega_M\delta(T)\int DX^{\mu}e^{-I_m }
\ee
where 
\be
I_m 
={1\over 4\pi} \int d^{2|2}z\, E\, {\cal D}_+X^{\mu}{\cal D}_-X^{\mu}\, ,
\qquad \quad
E \equiv \sdet E_M{}^A
\ee
Here, $X^{\mu}=x^{\mu}+\theta\psi_+^{\mu} +\bar\theta\psi_-^{\mu} + i
\theta \bar \theta F^\mu$, $\mu = 0,1,\cdots,9$ are scalar superfields,
and $\delta(T)$ indicates the torsion constraints.\footnote{Here as well
as in the superghost fields $B,C$ below, we shall henceforth omit
auxiliary fields such as $F^\mu$, since they can be conveniently
integrated out and do not play a significant role \cite{dp88, dp89}.} The
theory is invariant under sDiff(M), super Weyl, and sU(1) local gauge
transformations. The space of supergeometries $E_M{}^A$, $\Omega _M$,
modulo these symmetries is supermoduli space. In genus $h\geq 2$, it has
dimension $(3h-3|2h-2)$. Let $m^A=(m^a|\zeta ^ {\alpha})$,
$a=1,\cdots,3h-3$, $\alpha = 1,\cdots ,2h-2$, be parameters for a local 
slice  ${\cal S}$  transversal to the orbits of the gauge group. The
starting point for our considerations is the gauge-fixed
expression for ${\bf A}[\delta]$, which is given by \cite{dp88}, p. 967,
\be
\label{smeas}
{\bf A}[\delta]
=
\int |\prod_A dm ^A|^2 \int D(B\bar BC\bar CX^{\mu})
|\prod_A \delta(\<H_A|B\>)|^2  e^{-I_m -I_{gh}}
\ee
The ghost superfields $B=\beta+\theta b$, $C=c+\theta\gamma$ have U(1)
weights $3/2$ and $-1$ respectively, with action 
\be
I_{gh}={1\over 2\pi}\int \! d^{2|2}z\, E
\biggl ( B{\cal D}_-C + \bar B {\cal D} _ + \bar C \biggr )
\ee 
and $H_A\equiv (H_A)_ -{}^z$ are the Beltrami
superdifferentials tangent to the gauge slice
\be
(H_A)_- {}^z \equiv (-)^{A(M+1)} E_- {}^M {\p E_M {}^z\over\p m^A}
\ee 
The vertex operators for NS-NS states do not involve superghost fields,
and it suffices to consider the gauge-fixed measure in (\ref{smeas}) for 
the scattering of these states. However, we would like to stress that
(\ref{smeas})  is only a preliminary first step in the
construction of the desired superstring measure,
since it is a {\it non-chiral} measure on {\it supermoduli} space.

\section{Chiral Splitting}
\setcounter{equation}{0}

A first step in the construction of the superstring measure
is to extract from (\ref{smeas}) a chiral contribution.
In Wess-Zumino gauge, the supergeometry
$E_m{}^a=e_m{}^a+\theta\gamma^a\chi_m-{i\over 2}\theta\bar\theta e_m{}^aA$
decomposes into a zweibein $e_m{}^a$, a gravitino field 
$\chi_m{}^{\alpha}$ and an auxiliary field $A$.  The expression
(\ref{smeas}) as well as vertex operators $V_i(k_i, \epsilon _i)$ mix
fields of opposite chiralities such as $\chi_{\bar z}{}^+$ and
$\chi_z{}^-$, as can be seen in the components expression of $I_m$,
\bea
I_m 
&=&
{1\over 4\pi}
\int d^2z \biggl ( \partial_z x ^\mu \partial_{\bar z} x ^\mu
-\psi_+ ^\mu \partial_{\bar z}\psi_+ ^\mu
-\psi_- ^\mu \partial_z\psi_- ^\mu 
 \\
&& \hskip .75in
+\chi_{\bar z}{}^+\psi_+ ^\mu \partial_z x ^\mu
+\chi_z{}^-\psi_- ^\mu \partial_{\bar z} x ^\mu
-{1\over 2}\chi_{\bar z}{}^+\chi_z{}^-\psi_+ ^\mu 
\psi_- ^\mu \biggr )
\nonumber
\eea
They also involve the scalar fields $x^{\mu}(z)$ whose oscillator modes
are split, but whose momentum zero modes are not split since left and
right momenta coincide. As shown in \cite{dp89}, to obtain the chiral
amplitudes, we have to introduce internal loop momenta $p_I^{\mu}$,
$ I=1,\cdots,h$, $\mu=0,1,\cdots, 9$, 
and require the following effective
prescription for the scalar superfield correlation functions,
\be
\label{chispl}
\<\prod_{i=1}^NV_i(k_i, \epsilon _i )\>_{X^{\mu}}
=
\int dp_I^{\mu}\ \bigg | \bigg \< \prod_{i=1}^N
V_{i}^{chi}(k_i, \epsilon _i;p_I^{\mu})
\bigg \>_+ \bigg | ^2
\ee
Here, $\<\cdots \>_+$ denotes the fact that the effective rules for the
contractions of the vertex operators
$V_{i}^{chi}(k_i, \epsilon _i ;p_I^{\mu})$ are used, as given in Table 1.
\begin{table}[h]
\begin{center}
\begin{tabular}{|c||c|c|} \hline 
& {\rm Original} & {\rm Effective Chiral} 
                \\ \hline \hline
              {\rm Bosons}  
            & $x^{\mu}(z)$ 
            & $x_+^{\mu}(z)$        
             
 \\ \hline
              Fermions  
            & $\psi_+ ^\mu (z)$ 
            & $\psi_+ ^\mu (z)$
\\ \hline
 Internal Loop momenta
& None
& ${\rm exp}(p_I^{\mu}\oint_{B_I}dz\partial_zx^{\mu} _+)$       
            
 \\ \hline
              $x$-propagator  
            & $\<x^{\mu}(z)x^{\nu}(w)\>$ 
            & $-\delta^{\mu\nu}{\rm ln}\,E(z,w)$        
 \\ \hline
              $\psi_+$-propagator  
            & $\<\psi_+ ^\mu (z)\psi_+^{\nu}(w)\>$ 
            & $- \delta^{\mu\nu}S_{\delta}(z,w)$        
 \\ \hline
              Covariant Derivatives  
            & ${\cal D}_+$ 
            & $\partial_{\theta}+\theta\partial_z$        
 
 \\ \hline
\end{tabular}
\end{center}
\caption{Effective Rules for Chiral Splitting}
\label{table:1}
\end{table}
In this table, we have chosen a canonical homology $A_I,B_I$, 
$I=1,\cdots,h$ with canonical intersections $\#(A_I\cap
B_J)=\delta_{IJ}$, $E(z,w)$ is the prime form, and $S_\delta (z,w)$ is
the Szeg\"o kernel. The point of the effective rules is that they only
involve meromorphic notions, unlike the $x$-propagator
$\<x^{\mu}(z)x^{\nu}(w)\>$ which is given by the scalar Green's function
$\delta ^{\mu\nu}G(z,w)$. The superghost
correlation functions are manifestly split. For the superstring measure
alone, we obtain the following formula,
\be
\label{smeaschi}
{\bf A}[\delta] 
=
\int |\prod_A dm^A|^2
\int dp_I^{\mu}\
\bigg | e^{i\pi p_I^{\mu}\hat\Omega_{IJ}p_J^{\mu}}\,
{\cal A} [\delta]\, \bigg |^2\, 
=
\int \ {
\big | \ \prod_A dm^A\ {\cal A} [\delta]\,\big|^2 \over 
(\det\,{\rm Im}\,\hat \Omega)^5}\, 
\ee
where ${\cal A}[\delta]$ is the following effective chiral correlator
\be
{\cal A} [\delta] 
=
\bigg \< 
\prod_A\delta(\<H_A|B\>)
\exp \biggl \{ \int \! {d^2\! z \over 2 \pi} \chi_{\bar z}{}^+
S(z)\biggr  \}\bigg \> _+
\ee
$S(z)$ is the total supercurrent
\be
S(z)
=-{1\over 2}\psi_+^{\mu}\p_zx_+^{\mu}
+{1\over 2}b\gamma-{3\over 2}\beta\p_zc
-(\p_z\beta)c,
\ee
and $\hat\Omega_{IJ}$ is the super period matrix, defined by
\cite{dp88,dp89}
\be
\label{sper}
\hat\Omega_{IJ}
=
\Omega_{IJ}-{i\over 8\pi}\int \! d^2 \! z \int \! d^2 \! w
\ \omega_I(z)\chiz\hat S_{\delta}(z,w) \chiw\omega_J(w)
\ee
Here $\Omega_{IJ}$ is the period matrix corresponding to the
complex structure of the metric $g_{mn}=e_m{}^a e_n{}^b\delta_{ab}$
in the homology basis $\{A_I,B_I, \ I=1,\cdots ,h\}$;
$\{\omega_I(z), \ I=1,\cdots ,h\}$, is a basis of holomorphic Abelian
differentials dual to the $A_I$-cycles;
and $\hat S_{\delta}(z,w)$ is a modified Dirac propagator defined by
\be
\p_{\bar z}\hat S_{\delta}(z,w)
+{1\over 8\pi}\chiz\int d^2x\ \p_z\p_x\ln E(z,x) \chix \hat S_{\delta}(x,w)
=2\pi\delta(z,w).
\ee
The chirally split expression (\ref{smeaschi}) is our first significant
departure from the proposals of other authors in the late 1980's, in that
it is the super period matrix $\hat\Omega_{IJ}$ which appears as
covariance of the internal loop momenta $p_I^{\mu}$, and not the period
matrix $\Omega_{IJ}$. More important, we observe that a correct chiral
splitting points then to the super period matrix $\hat\Omega_{IJ}$ as the
proper locally supersymmetric moduli for gauge-fixing.

\section{Local Supersymmetry and Gauge-Fixing}
\setcounter{equation}{0}

The expression for ${\bf A}[\delta]$ in (\ref{smeas}) and (\ref{smeaschi})
is an integral  over supermoduli space.
The main problem in superstring perturbation theory is how to integrate
out the odd supermoduli $\zeta ^ {\alpha}$ in (\ref{smeas}) and
(\ref{smeaschi}) to reduce
${\bf A}[\delta]$ to a measure $d\mu[\delta]$ over moduli space
\be
\label{mu}
d\mu[\delta]
=\prod_{a=1}^{3h-3}dm^a\, \int \prod_{\alpha=1}^{2h-2}
d\zeta^{\alpha}
\ {\cal A}[\delta]
\ee
Let ${\cal S}$ be a gauge slice for supermoduli, obtained by choosing a
$3h-3$ dimensional gauge slice of zweibeins $e_m{}^a$, a $2h-2$
dimensional slice of gravitino sections $\chi_{\alpha}$, and setting
$\chi=\sum_{\alpha=1}^{2h-2} \zeta^ {\alpha} \chi_{\alpha}$. Naively, it
may seem that the natural way of descending from supermoduli space to
moduli space is to use the projection 
\be
\label{bproj}
E_M{}^A\longrightarrow e_m{}^a
\ee
This has been the method followed in the literature on superstring
perturbation theory, but it has resulted in amplitudes which depend on
the gauge slice  ${\cal S}$  chosen. The origin of this apparent
ambiguity is the fact that the projection (\ref{bproj}) is {\sl not}
invariant under local worldsheet supersymmetry. The remedy, originally
proposed in \cite{dp88} and carried out in the  present paper, is to use
instead the projection
\be
\label{proj}
E_M{}^A\longrightarrow \hat\Omega_{IJ}
\ee
where $\hat\Omega_{IJ}$ is the super period matrix defined by
(\ref{sper}).
The correct moduli measure is
obtained from the supermoduli measure by integrating along the fibers
of (\ref{proj}). For genus 2, this is implemented by
choosing $\{m^A\}_{A=1,2,3}=\{\hat\Omega_{IJ}\}_{1\leq I\leq J\leq 2}$
(instead of $\{\Omega_{IJ}\}_{1\leq I\leq J\leq 2}$),
and integrating in $\zeta^{\alpha}$, $\alpha =1, 2$.
We describe next some important steps in this process.

\section{The Moduli Space Measure}
\setcounter{equation}{0}

The change of projection from (\ref{bproj}) to (\ref{proj}) leads to three
modifications which require particular care. First, the  Beltrami
superdifferentials get modified to superdifferentials
$H_A=\bar\theta(\mu_A-\theta \nu_A)$ (in Wess-Zumino gauge) with {\it
both} components $\mu_A$ and $\nu_A$ usually non-zero. Second, the
correlation functions in (\ref{smeas}) and (\ref{smeaschi}) were
originally given in the  metric $g_{mn}$ corresponding to $\Omega_{IJ}$.
They have now to be  re-expressed in a new metric $\hat g_{mn}$
corresponding to $\hat\Omega_{IJ}$. This deformation of metrics 
requires an insertion of the stress tensor integrated against a
Beltrami differential $\hat \mu$ that represents this change of metrics.
Third, we note that for given $\hat\Omega_{IJ}$, the metric $\hat g_{mn}$
is not unique. Thus the choice of $\hat g_{mn}$, or equivalently
$\hat\mu$, should be viewed as an additional gauge choice, of which the
final amplitude has also to be shown to be independent.

\medskip

Due to the complicated nature of the Beltrami superdifferentials $H_A$,
the superghost  correlation functions in the presence of $\delta
(\<H_A|B\>)$ are not convenient meromorphic objects. To
circumvent this problem, we change basis to super-Beltrami differentials
$H_a^*(z,\theta)=\bar\theta\,\delta(z,p_a)$ and 
$H_\alpha^*(z,\theta)
=\bar\theta\theta\,\delta(z,q_{\alpha})$ which are 
$\delta$-functions at points
$p_a$ and $q_\alpha$ respectively. Assuming that correlation functions
are considered with superghost field $B$-independent vertex operators
only (as is always the case for NS states \cite{dpvertex}), $H_A$ is
effectively integrated versus superholomorphic $B$'s. A change of basis
for $H_A$ is then carried out and produces an associated Jacobian, 
\be
\prod _A \delta (\< H_A | B \>)
=
{\sdet \<H_A |\Phi _C \> \over \sdet \< H_A^* | \Phi _C\>}
\prod _a b(p_a) \prod _\alpha \delta (\beta(q_\alpha))
\ee
for an arbitrary basis of superholomorphic 3/2 forms $\Phi _C$. By
construction, all dependence on $p_a$ and $q_\alpha$ cancels out in the
full measure and amplitude. Upon choosing $\Phi_C$
to satisfy $\<H_A | \Phi _C\>=\delta _{AC}$, we find the
following expression for the chiral measure,
\bea
{\cal A} [\delta]
& = & 
{\< \prod _a b(p_a) \prod _\alpha \delta (\beta (q_\alpha)) \>
\over \det \Phi _{IJ+} (p_a)  \det \< H_\alpha | \Phi ^* _\beta\>}
\biggl \{ 1 - {1 \over 8 \pi ^2} \int \! d^2z \chiz \int \! d^2 w
\chiw
\< S(z) S(w)\> 
\nonumber \\
&& \hskip 2in+ {1 \over 2 \pi} \int d^2 z \hat \mu _{\bar z }{}^z \<
T(z)\> \biggr \}
\eea
Here, $\Phi_{IJ}$ are the odd superholomorphic 3/2 differentials
corresponding to the covectors $d\hat \Omega _{IJ}$ on supermoduli space,
$\Phi _{IJ+}$ is the $\theta$ component of 
$\Phi_{IJ}=\theta\Phi_{IJ+}+\Phi_{IJ0}$,  and $\Phi ^* _\beta
=\theta\Phi^*_{\beta+}+\Phi^*_{\beta0}$ is the basis for the even
superholomorphic 3/2 differentials normalized by $\Phi ^* _{\beta 0} 
(q_\alpha) = \delta _{\alpha \beta}$ and $\Phi ^* _{\beta +} (p_a) =0$.

\medskip

Taking all this into account, and expanding the 
finite-dimensional determinants in powers of $\chi$, we arrive at the
following formula
\be
\label{finamp}
{\cal A}[\delta] 
=
i \ {\< \prod _a b(p_a) \prod _\alpha \delta (\beta (q_\alpha)) \>
\over \det \bigl (\omega _I \omega _J (p_a) \bigr )
 \cdot \det \< \chi _\alpha | \psi ^* _\beta\>}
\biggl \{ 1  + {\cal X}_1 + {\cal X}_2 + {\cal X}_3 + {\cal X}_4 +   
{\cal X}_5 + {\cal X}_6 \biggr \}
\ee
where all correlation functions are now written with respect to the
$\hat\Omega_{IJ}$ complex structure.\footnote{Henceforth, the original
$\Omega_{IJ}$ will no longer enter, and to simplify notations we denote
$\hat\Omega_{IJ}$ by $\Omega_{IJ}$.}  
The various terms $\{{\cal X}_i\}_{i=1}^6$ in (\ref{finamp})
have the following origins. The term ${\cal X}_1$ is the familiar
contribution arising from two supercurrent insertions.
All the other terms are more subtle and incorporate the effect
of using $\hat\Omega_{IJ}$ as supermoduli invariant. The term $\X_2$
arises from the stress tensor insertion; the term $\X_3$ arises when
passing from the metric $g_{mn}$ to the
metric $\hat g_{mn}$ in  $\det \Phi _{IJ+}(p_a)$
and  $\det \<H_\alpha |\Phi _\beta ^*\>$; the terms $\X_4$  and
$\X_5$ arise from the remaining $\chi$-dependence of
$\det \Phi _{IJ+}(p_a)$; and the term $\X_6$ arises from the remaining
$\chi$-dependence of $\det \<H_\alpha | \Phi ^* _\beta\>$. The term $\X_2
+ \X_3$ thus contains all the effects of passing from the metric $g_{mn}$
to the metric $\hat g_{mn}$ via the Beltrami differential $\hat \mu$.
Using its expression below in terms of 
the holomorphic differential $T^{IJ} \omega _I (z) \omega _J (w)$, it 
will be manifest that  the measure depends only on the moduli of $\hat
g_{mn}$ and not on the slice chosen.

\medskip

More specifically, the quantities $\psi_{\beta}^*$ are the holomorphic
$3/2$ differentials normalized at the points $q_{\alpha}$ by $\psi ^*
_\beta (q_\alpha)=\delta _\beta {}^\alpha$, and the Green's functions
$G_2(z,w)$ and $G_{3/2}(z,w)$ are of tensor type $(2,-1)$ and $(3/2,-1/2)$
respectively in $z$ and $w$, and normalized so that $G_2(p_a,w)=0$ and
$G_{3/2} (q_\alpha, w)=0$. The terms $ \X_i, \ i=1,\cdots ,6$ are
then defined as follows, 
\bea
\label{Xes}
{\cal X}_1
&=&
 - {1 \over 8 \pi ^2} \int \! d^2z \chiz \int \! d^2 w \chiw \< S(z) S(w)\> 
\nonumber \\
{\cal X}_2
+
{\cal X}_3
&=&
+{1\over 16\pi^2}
\int d^2z\int d^2w\,\chiz\chiw\, T^{IJ} 
\omega_I(z) S_{\delta}(z,w) \omega_J(w)
\nonumber\\
{\cal X} _4 
&=& + {1 \over 8\pi ^2} \int \! d^2w \ \p _{p_a} \pw \ln E(p_a,w) \chiw 
\int \! d^2u S_\delta (w,u) \chiu \varpi ^* _a(u)
\nonumber \\
{\cal X} _5 
&=& + {1 \over 16 \pi ^2} \int \! d^2u \int \! d^2v S_\delta (p_a,u) \chiu  
\p _{p_a} S_\delta (p_a,v) \chiv \varpi _a  (u,v) 
\nonumber \\
{\cal X} _6
&=& + {1 \over 16 \pi ^2} \int \! d^2z \chi _\alpha ^* (z) \int \!
d^2w G_{3/2} (z,w) \chiw \int \! d^2 v \chiv \Lambda _\alpha (w,v)  
\eea
The sections $\chi_{\beta}^*(z)$ are the linear combinations
of the $\chi_{\alpha}(z)$ normalized by
$\<\chi_{\beta}^*|\psi_{\alpha}^*\> =\delta_{\beta\alpha}$
and $T^{IJ}\omega_I(w)\omega_J(w)$
is the holomorphic quadratic differential defined by
\bea
T^{IJ} \omega_I\omega_J(w)
&=&
\< T(w) \prod_{a=1}^3 b(p_a) \prod_{\alpha=1}^2
\delta(\beta(q_{\alpha}))\> \bigg / \< \prod_{a=1}^3 b(p_a)
\prod_{\alpha=1}^2 \delta(\beta(q_{\alpha}))\>
 \\
&& -2 \sum_{a=1}^3 \p_{p_a} \p_w\ln\,E(p_a,w) \varpi^* _a (w)
\nonumber\\
&&
+\int d^2z\chi_{\alpha}^*(z) \biggl (
-{3\over 2}\p_wG_{3/2}(z,w)\psi_{\alpha}^*(w)
-{1\over 2}G_{3/2}(z,w)(\p\psi_{\alpha}^*)(w)
\nonumber\\
&& \hskip 1.1in 
+G_2(w,z)\p_z\psi_{\alpha}^*(z)
+{3\over 2}\p_zG_2(w,z)\psi_{\alpha}^*(z) \biggr )
\nonumber
\eea
Here, $T(z)$ is the total stress tensor
\be
\label{stress}
T(z)
=-{1\over 2}\p_z x^{\mu}\p_zx^{\mu} + \half \psi_+^{\mu}\p_z\psi_+^{\mu}
+c\p_zb -(\p_zc)b -2\beta\p_z\gamma
-{3\over 2}\gamma\p_z\beta +{3\over 2}(\p_zc)b
\ee
and $\Lambda_{\alpha}$ is defined by
\be
\Lambda_{\alpha}(w,v)
=
2G_2(w,v)\p_v\psi_{\alpha}^*+3\p_vG_2(w,v)\psi_{\alpha}^*(v)
\ee
Finally, $\varpi^* _a$ and $\varpi_a$ are holomorphic 1-forms in $u$
and $v$ defined by $\varpi^* _a (u) = \varpi _a (u,p_a)$ and the
determinants of $3\times 3$ matrices
\bea
\varpi _a (u,v) 
& = & 
{\det \{ \omega _I \omega _J(p_b [ u,v;a ] ) \}
\over \det \{ \omega _I \omega _J (p_b) \} }
\nonumber \\
&& \nonumber \\
\omega _I \omega _J (p_b [ u,v; a ] ) 
&=& \left \{ \matrix{
\omega _I \omega _J(p_b) & \qquad b\not= a\cr
\half (\omega _I(u)\omega _J(v) + \omega _I(v) \omega _J(u) )  &
\qquad b=a
\cr}
\right .
\eea

The above gauge-fixed amplitude (\ref{finamp}) is independent of the 
points  $p_a$  and $q_{\alpha}$. Furthermore, it satisfies the crucial
requirement of invariance under infinitesimal deformations
of the gauge slice ${\cal S}$, produced by $\delta_{\xi} (\chi _\alpha )
_{\bar z}{}^+=-2\p_{\bar z}\xi_{\alpha}^+$,
\be
\delta_{\xi}\bigg(
\int \prod_{\alpha=1}^2d\zeta^{\alpha}
\ {\cal A}[\delta]\bigg)
=0
\ee
with $\xi_{\alpha}^+$ two arbitrary spinor fields, generators of local
supersymmetry transformations.

\section{Slice Independence and Absence of Ambiguities}
\setcounter{equation}{0}

We specialize now to gauge slices ${\cal S}$ given by $\delta$ functions
\be
\chiz=\zeta ^1 \delta (z,x_1) + \zeta ^2 \delta (z,x_2)
\ee
where $x_\alpha,\ \alpha =1,2$ are two arbitrary fixed points on $\Sigma$.
It may be shown explicitly that for each $\Omega _{IJ}$,
(\ref{finamp}) is a holomorphic scalar in  all points $p_a$, $q_\alpha$
and $x_\alpha$, and thus independent thereof.  As a result, the
gauge-fixed chiral measure (\ref{finamp}) is not just invariant under
infinitesimal deformations of gauge slices, but more globally,
independent of the gauge slices ${\cal S}$ themselves, at least when they
are $\delta$-functions. 

It is natural to let $q_{\alpha}$ coincide with $x_{\alpha}$.
In this limit, the positions of the supercurrent insertions $S(x_\alpha)$
tend to those of the superghost insertion $\delta (\beta (q_\alpha))$.
In the early literature on superstring perturbation theory
\cite{vv}, the picture changing operator $Y(z)$ had been naively
identified with $ Y(z) = \delta(\beta(z))S(z)$, but the difficulties
inherent to taking this product at coincident points had been obscure.
Thanks to the expression (\ref{finamp}) and (\ref{Xes}), we see now that
these difficulties cannot be ignored.  Indeed, the corresponding
term $\X_1$ fails to admit a limit as $x_\alpha \to q_\alpha$,
and the correct limit is more subtle: it requires the contribution of the  
finite-dimensional determinant $\det \< H_\alpha | \Phi ^* _\beta\>$ in
$\X_6$, and it is only the sum $\X_1 +
\X_6$ which admits a finite limit.

\medskip

With the subtleties taken into account, 
the gauge-fixed amplitude ${\cal A}[\delta]$
can be expressed in the form
(\ref{finamp}), with the following simpler expressions for the terms
$\X_i$, $i=1,\cdots ,6$,
\bea
\label{q=x}
\X_1 + \X _6 &=&
{\zeta ^1 \zeta ^2 \over 16 \pi ^2}
\biggl [
-10 \ S_\delta (q_1,q_2) \p _{q_1} \p _{q_2} \ln E(q_1,q_2)
\nonumber \\
&&
\qquad -\p_{q_1}G_2(q_1,q_2)\p\psi_1^*(q_2)
+\p_{q_2}G_2(q_2,q_1)\p\psi_2^*(q_1)\nonumber\\
&&
\qquad 2G_2(q_1,q_2)\p\psi_1^*(q_2)f_{3/2}^{(1)}(q_2)
-2G_2(q_2,q_1)\p\psi_2^*(q_1)f_{3/2}^{(2)}(q_1)\biggr]
\nonumber\\
\X _2 &=&
{\zeta ^1 \zeta ^2 \over 16 \pi ^2}
\omega _I(q_1) \omega _J(q_2) S_\delta (q_1,q_2) 
\biggl [  \p _I \p _J \ln {\tet [\delta ](0)^5 \over \tet
[\delta ](D_\beta )} +   \p _I \p _J \ln \tet (D_b ) \biggr ]
\nonumber \\
\X _3 &=&
{\zeta ^1 \zeta ^2 \over 8 \pi ^2}
S_\delta (q_1,q_2) \sum _a  \varpi _a (q_1, q_2) \biggl [ B_2(p_a) +
B_{3/2}(p_a) \biggr ]
 \\
\X_4 &=& {\zeta ^1 \zeta ^2 \over 8 \pi ^2}
S_\delta (q_1,q_2) \sum _a \biggl [
\p _{p_a} \p _{q_1} \ln E(p_a,q_1) \varpi ^* _a(q_2)
+ \p _{p_a} \p _{q_2} \ln E(p_a,q_2) \varpi ^* _a(q_1) \biggr ]
\nonumber \\
\X _5 &=&
{\zeta ^1 \zeta ^2 \over 16 \pi ^2}
\sum _a \biggl [ 
S_\delta (p_a, q_1) \p _{p_a} S_\delta (p_a,q_2) 
- S_\delta (p_a, q_2) \p _{p_a} S_\delta (p_a,q_1) \biggr ] 
\varpi _a  (q_1,q_2) \, .
\nonumber
\eea
Here 
$D_b=\sum_{a=1}^3p_a-3\Delta$,
$D_{\beta}
= \sum_{\alpha=1}^2 q_{\alpha}-2\Delta$; 
the terms $B_{3/2}$ and $B_2$ are given by
\bea
\label{bthreehalfs}
B_2(w) &=& -27 T_1(w) +
 \half f_2 (w)^2 -{3 \over 2} \pw f_2(w)-2 \sum _a \p_{p_a} \pw \ln
E(p_a,w) \varpi ^* _a (w) \, .
\nonumber\\
B_{3/2}(w) &=& 12 T_1(w) 
-\half f_{3/2}(w)^2 + \pw f_{3/2}(w) 
 \\
&& +\int \! d^2z \chi ^* _\alpha (z) \biggl (
-{3 \over 2} \pw G_{3/2} (z,w) \psi ^* _\alpha (w)
  -\half  G_{3/2} (z,w) \pw \psi ^* _\alpha (w)
\nonumber \\
&& \qquad \qquad \qquad \qquad 
+ G_2 (w,z) \pz \psi ^* _\alpha (z) + {3 \over 2} \pz G_2 (w,z) \psi
^* _\alpha (z) \biggr );
\nonumber
\eea
the expression $-T_1(z)$ is the vev of the chiral scalar boson stress 
tensor defined by $E(z,w)=(z-w)+(z-w)^3T_1(w)+{\cal O}(z-w)^4$.
The expressions $f_{3/2}(w)$, $f_2(w)$ are given by
\be
f_n(w)
=
\omega_I(w)\p_I{\rm ln}\,\tet [\delta](D_n)
+\p_w{\rm ln}\,\bigg ( \sigma(w)^{2n-1}\prod_{i=1}^{2n-1}E(w,z_i)\bigg )
\ee
with $\{z_i\}=\{p_a\}$ for $n=2$, and $\{z_i\}=\{q_{\alpha}\}$
for $n=3/2$, and
\bea
f_{3/2}^{(1)}
(q_2)&=&
\omega_I(q_1)\p_I\ln\,\tet [\delta](q_2-q_1+D_{\beta})
+
\p_{q_1}\ln\big(E(q_1,q_2)^2\sigma(q_1)^2)\nonumber\\
f_{3/2}^{(2)}
(q_1)&=&
\omega_I(q_2)\p_I\ln\,\tet [\delta](q_1-q_2+D_{\beta})
+
\p_{q_2}\ln\big(E(q_2,q_1)^2\sigma(q_2)^2).
\eea 

The corresponding chiral measure $d\mu[\delta](\Omega)$ is independent 
of $p_a$ and $q_{\alpha}$. This is of course a consequence of the
independence of the earlier formulas from $p_a,q_{\alpha}$ and 
$x_{\alpha}$. But it can also be verified directly from (\ref{q=x}).
This provides a direct proof
of the absence of ambiguities in the gauge-fixed
expression measure $d\mu[\delta](\Omega)$.

\section{The Measure as a Modular Form}
\setcounter{equation}{0}

It remains to express the moduli measure in terms of modular forms.
In genus $h=2$, there are $10$ even spin structures, denoted $\delta$,
and $6$ odd spin structures, denoted $\nu$. Each even spin structure
$\delta$ can be written as $\delta=\nu_1+\nu_2+\nu_3$,
where the $\nu_i$'s are odd and pairwise distinct.
The mapping $\{\nu_1,\nu_2,\nu_3\}\to\delta$ is $2$ to $1$,
with $\{\nu_1,\nu_2,\nu_3\}$ and 
$\{\nu_1,\cdots,\nu_6\}\setminus\{\nu_1,\nu_2,\nu_3\}$
corresponding to the same even spin structure.
In the hyperelliptic representation, the surface $\Sigma$
is given by $s^2=\prod_{k=1}^6(x-u_k)$ with 6 branch points
$u_k$. Each odd spin structure $\nu$ corresponds to
a unique branch point $u_{\nu}$. Each even spin structure
$\delta=\nu_1+\nu_2+\nu_3$ corresponds then to a
partition of the 6 branch points into two sets
of 3 branch points each, namely
$\{u_{\nu_1},u_{\nu_2},u_{\nu_3}\}$ and
$\{u_k\}_{k=1}^6\setminus \{u_{\nu_1},u_{\nu_2},u_{\nu_3}\}$.

\medskip

Define $\Xi_6[\delta](\Omega)$ by the following combination of 
$\tet$-constants, $\tet [\delta] (\Omega) \equiv \tet [\delta](0,\Omega)$,
\be
\Xi_6[\delta] (\Omega) \equiv \sum _{1\leq i < j \leq 3}
\<\nu_i |\nu _j\> \prod _{k=4,5,6} \tet [\nu_i + \nu_j +\nu_k]^4 (\Omega)
\ee
which depends only on $\delta=\nu_1+\nu_2+\nu_3$.
Here $\<\nu_i|\nu_j\>={\rm exp}
4\pi i(\nu_i'\nu_j''-\nu_j'\nu_i'')$ is the relative signature of the spin 
structures $\nu_i=(\nu_i'|\nu_i'')$, $\nu_j=(\nu_j'|\nu_j'')$.
We shall also need the well-known modular form of weight 10, defined by
$\Psi _{10} (\Omega) \equiv \prod _\delta \tet [\delta ](\Omega )^2$. The
modular transformation properties of the spin structures are given by
\cite{igusa},
\be
\left (\matrix{ \tilde \delta' \cr \tilde \delta ''\cr}  \right )
=
\left ( \matrix{D & -C \cr -B & A \cr} \right )
\left ( \matrix{ \delta ' \cr \delta '' \cr} \right )
+ \half \ {\rm diag} 
\left ( \matrix{CD^T  \cr AB^T \cr} \right )\, ,
\qquad
\left ( \matrix{A & B \cr C & D \cr} \right ) \in Sp(4,{\bf Z})
\ee
while the period matrix transforms as $\tilde \Omega = (A\Omega +B)
(C\Omega +D)^{-1}$, $\det\,\Im\tilde\Omega
=|\det\,(C\Omega+D)|^{-2}
(\det\,\Im\Omega)$, and the $\tet$ constants obey 
\bea
\tet [\tilde \delta ] ^4 (\tilde \Omega) 
& = & 
\epsilon ^4 \det (C\Omega + D)^2 \tet [\delta ] ^4 (\Omega)
\nonumber \\
\Xi _6 [\tilde \delta ] (\tilde \Omega) 
& = & 
\epsilon ^4 \det (C\Omega + D)^6 \Xi _6 [\delta ] (\Omega )
\nonumber \\
\Psi _{10} (\tilde \Omega) 
& = & \det (C\Omega + D)^{10} \Psi _{10}(\Omega)
\eea
Here, $\epsilon$ depends on both $\delta $ and the modular transformation,
and satisfies $\epsilon ^8=1$.\footnote{It is important to remark that
$\Xi_6[\delta](\Omega)$ is not a modular form since it depends on $\delta$
and the sign factor $\epsilon^4$ arises in its transformation laws. $\Xi
_6 [\delta](\Omega)$ is thus different from the modular form
$\Psi_6(\Omega)$ of weight 6, which is obtained by summing 60 (syzygous)
products of three $\tet ^4$ each, see \cite{igusa}.}

\medskip
 
Recall that the gauge-fixed amplitude (\ref{finamp}), (\ref{q=x}) has 
been shown to be independent of the points $p_a$ and $q_{\alpha}$. We may
then choose the points $p_a$ to make up either one of the two sets of 3
branch points  defining $\delta$. It turns out that the most convenient
choice for $q_{\alpha}$ is to constrain them by the following {\sl split
gauge condition}
\be
\label{split}
S_{\delta}(q_1,q_2)=0
\ee
The latter choice implies  $\hat\Omega_{IJ}=\Omega_{IJ}$, although no
direct use is made of this fact. 

\medskip

With (\ref{q=x}), the above choice of $p_a$ and (\ref{split}), the chiral
amplitude (\ref{finamp}) for the  spin structure $\delta$  can be
evaluated explicitly. We obtain in this way one of the main results of
this paper, which gives the contribution of each even spin structure
$\delta$ to the superstring measure
\be
d\mu[\delta] (\Omega)
=
\prod_{I\leq J}d\Omega_{IJ} \int \prod _\alpha d\zeta^\alpha {\cal A}
[\delta] =
{1\over 16\pi^6} \prod_{I\leq J}d\Omega_{IJ}
\,
{\Xi_6[\delta](\Omega)\tet[\delta]^4(\Omega)\over
\Psi_{10}(\Omega)} 
\ee
Using the modular transformation properties of the measure $\prod
_{I\leq J}d\tilde\Omega _{IJ}
=\det\,(C\Omega+D)^{-3}\prod_{I\leq J}d\Omega_{IJ}$, and the above modular transformation rules, 
we obtain as an immediate consequence that the measure $d\mu[\delta]$
is modular covariant, i.e.,
\be
\label{gso}
d\mu [\tilde \delta] (\tilde \Omega) = 
\det\,(C\Omega+D)^{-5}d\mu [\delta ](\Omega),
\ee
without any multiplicative phase factors arising.

\section{The GSO Projection and Cosmological Constant}
\setcounter{equation}{0}

To implement the GSO projection \cite{sw}, we have to sum over spin
structures. Given the above modular transformation laws of the measure,
there is a unique choice of relative phase factors (namely all $\eta
_\delta = 1 $) leading to a modular form, 
\be
\label{measuremu}
d\mu(\Omega)
\equiv\sum_{\delta}d\mu[\delta](\Omega).
\ee
In genus 1, GSO phases were related to the sign factors arising in the
unique genus 1 Riemann relation. In genus 2, however, there is a different
Riemann relation for each of the 6 odd spin structures $\nu$,
\be
\sum_{\delta}\<\nu|\delta\>\tet^4[\delta](\Omega)=0, 
\ee
and there is neither a unique nor a natural choice that leads to modular
invariance. Therefore, the uniqueness and naturality of the relative
phases in (\ref{measuremu}) (and hence of the GSO projection in the even
spin structure sector) should be viewed as a major advantage over any
mechanism for modular invariance based on Riemann identities.
 
\medskip

Since the right hand side of (\ref{measuremu}) is now known to be a
modular form, it can be shown to vanish identically in $\Omega$,
\be
\label{vanishing}
\sum_{\delta}\Xi_6[\delta](\Omega)\tet ^4[\delta](\Omega)=0
\ee
by examining its behavior along the divisor of Riemann surfaces with
nodes. This identity does not follow from the Riemann identities. Rather,
it is equivalent to the genus 2 identity that a modular
form of weight 8 must be proportional to the square of the unique modular
form of weight 4. Altogether, we have obtained a proof from
first principles that the two-loop cosmological constant $\Lambda$
is given by an integral over moduli space which vanishes
point by point, and hence $\Lambda$ must also vanish. For the Type II
superstrings,
\be
\label{summation}
\Lambda
={1\over 2^8\pi^{12}}\int
\ \big(\det\,\Im\,\Omega\big)^{-5} 
\biggl |\
\prod_{I\leq J}d\Omega_{IJ}\ 
{\sum_{\delta}\Xi_6[\delta](\Omega)\tet ^4[\delta](\Omega)
\over \Psi_{10}(\Omega)}\ \biggr |^2 
=0
\ee
We have an analogous expression for the heterotic strings
which also vanishes using (\ref{vanishing}). Finally, the asymptotic
behavior of the measure as $\Omega$ approaches the boundary of moduli
space for a separating degeneration (the non-separating case is
analogous) is obtained by decomposing the period matrix and the spin
structure consistently with the separation,
\bea
\Omega = \left ( \matrix{\tau_1 & \tau \cr \tau & \tau_2} \right )
\qquad \qquad 
\delta = \left ( \matrix{\mu _1 \cr \mu _2} \right ) \quad {\rm or} \quad
\delta = \left ( \matrix{\nu_0 \cr \nu_0} \right )
\eea
Here, $\tau \to 0$ in the degeneration while the genus 1 moduli $\tau
_{1,2}$ remain finite; $\mu_1$  and $\mu_2$ are one of the three even and
$\nu _0$ is the unique odd spin structures on each separated genus 1
surface. The separating degeneration limit of the measure is then given by
\bea
d \mu \left [ \matrix{\mu _1 \cr \mu _2} \right ] (\Omega) 
& \to &
{1 \over 2^{10} \pi ^8 \tau ^2} \ 
\<\mu _1 | \nu _0 \> \< \mu _2 |\nu _0\> {\tet _1 [\mu_1 ] (\tau _1)^4
\tet _1 [\mu _2 ](\tau_2)^4 \over
\eta(\tau _1)^{12} \eta (\tau _2)^{12}} \ d\tau_1 \ d\tau_2 \ d\tau 
\nonumber \\
d \mu \left [ \matrix{\nu_0 \cr \nu_0} \right ] (\Omega) 
& \to &
 {3 \tau^2 \over 2^6 \pi ^4} \ d\tau_1 \ d\tau_2 \ d\tau 
\eea
The first 9 spin structures (with even spin structures $\mu_1$ and $\mu_2$
on each genus 1 part) exhibit the tachyon pole on each genus 1 part, and
the tachyon and massless intermediate state divergences, as
expected.\footnote{This behavior for fixed $\delta$ coincides with the
one of the bosonic string \cite{div,dp88}, as expected.} The last spin
structure (with odd spin  structure $\nu_0$ on each genus 1 part) has
no tachyon and no massless intermediate state divergences, as
expected.

\section{Scattering Amplitudes}
\setcounter{equation}{0}

The vertex operators for the scattering of $N$ massless bosons
are given by
\be
\prod_{i=1}^N V(k_i,\epsilon_i)
= \prod_{i=1}^N \int \! d^{2|2} z_i\, E(z_i)\,
\epsilon_i ^{\mu _i} \bar \epsilon_i ^ {\bar \mu _i}
{\cal D}_+X^{\mu_i} {\cal D}_-X^{\bar\mu_i}
e^{ik_{i\mu_i}X^{\mu_i}}(z_i)
\ee
As in the case of the measure, the superstring scattering amplitudes
require a GSO summation over spin structures of the conformal blocks of 
$\<\prod_{i=1}^NV(k_i,\epsilon_i)\>_X$ in the $X^{\mu}$ superconformal
field theory. The following formulas together with (\ref{vanishing})
are the proper analogues in genus 2 of the Riemann identities in genus 1,
and  may be used to carry out the required summations,
\bea
&& 
\sum _\delta \Xi _6 [\delta] (\Omega)  \tet [\delta ] (\Omega)^4
S_\delta (z_1,z_2) ^2 =0
\nonumber \\
&& 
\sum _\delta \Xi _6 [\delta] (\Omega)  \tet [\delta ] (\Omega) ^4 
S_\delta(z_1,z_2) S_\delta (z_2,z_3) S_\delta (z_3,z_1) =0
\eea
The 0-, 1-, 2- and 3-point functions in both the Type II and the
heterotic strings are then found to vanish pointwise on moduli space and
without the appearance of boundary terms.  

\medskip

The 4-point function receives contributions from two distinct parts. The
first arises from the connected part of the correlators
\bea
\label{connected}
\< S(z) S(w) \prod_{i=1}^4 V(k_i,\epsilon_i)^{chi} \>
\qquad {\rm and} \qquad
\< T(z) \prod_{i=1}^4 V(k_i,\epsilon_i)^{chi} \>\, .
\eea
The second arises from the disconnected part
\bea
\label{disconnected}
\< S(z) S(w) \> \< \prod_{i=1}^4 V(k_i,\epsilon_i)^{chi} \>
\qquad {\rm and} \qquad
\< T(z)\> \< \prod_{i=1}^4 V(k_i,\epsilon_i)^{chi} \>\, .
\eea
of these correlators and combines with the gauge fixing determinants into
a contribution proportional to the measure $d\mu [\delta](\Omega)$. The
connected part is more complicated and requires an independent treatment
to appear in a later publication. 

\medskip

The disconnected part (for example for the Type II superstrings) is given
by
\bea
\label{fourpoint}
\<\prod_{i=1}^4V(\epsilon_i,k_i)\>
=
g_s ^2 \delta (k) \int 
{\bigl | \prod_{I\leq J}d\Omega_{IJ} \bigr |^2 \over 
(\det \,\Im \Omega)^5} 
\prod_{i=1}^4 \int _\Sigma d^2z_i\, \bigl | {\cal F} \bigr |^2 
\exp \biggl (-\sum_{i<j}k_i \! \cdot \! k_jG(z_i,z_j)\biggr )
\eea
Here, $g_s$ is the string coupling, the scalar Green's function is given
by
\be
G(z,w) = -{\rm log}|E(z,w)|^2 + 2\pi \Im \int_z^w \!
\omega_I \ (\Im\,\Omega)^{-1}_{IJ} \Im \int_z^w \! \omega_J
\ee 
while $k$ is the total momentum, and $\F$ is a holomorphic 1-form in
each $z_i$, given by
\bea
\F = C_{{\cal S}}\, {\cal S} (1234) + \sum _{(i, j,k) = {\rm perm}(2,3,4)}
C_{{\cal T}} (1i|jk) \,{\cal T}(1i|jk) 
\eea
The combinations $C_{{\cal S}}$ and $C_{{\cal T}}$ are kinematical
factors, which depend only on the polarization vectors $\epsilon _i$ and
the external momenta $k_i$ through the gauge invariant combinations
$f_i ^{\mu \nu} \equiv \epsilon _i ^\mu k _i ^\nu - \epsilon _i ^\nu k_i
^\mu$ and are given by
\bea
C_{{\cal S}} &=&
f_1 ^{\mu \nu} f_2 ^{\nu \mu} f_3 ^{\rho \sigma} f_4 ^{\sigma \rho} +
f_1 ^{\mu \nu} f_2 ^{\rho \sigma} f_3 ^{\nu \mu} f_4 ^{\sigma \rho} +  
f_1 ^{\mu \nu} f_2 ^{\rho \sigma} f_3 ^{\sigma \rho} f_4 ^{\nu \mu}
\\ &&
- 4 f_1 ^{\mu \nu} f_2 ^{\nu \rho} f_3 ^{\rho \sigma} f_4 ^{\sigma \mu} 
- 4 f_1 ^{\mu \nu} f_2 ^{\rho \sigma} f_3 ^{\nu \rho} f_4 ^{\sigma \mu} 
- 4 f_1 ^{\mu \nu} f_2 ^{\nu \rho} f_3 ^{\sigma \mu} f_4 ^{\rho \sigma} 
\nonumber \\
C_{{\cal T}} (ij|kl) &=&
f_i ^{\mu \nu} f_j ^{\rho \sigma} f_k ^{\nu \mu} f_l ^{\sigma \rho} -
f_i ^{\mu \nu} f_j ^{\rho \sigma} f_k ^{\sigma \rho} f_l ^{\nu \mu} +2
f_i ^{\mu \nu} f_j ^{\nu \sigma} f_k ^{\sigma \rho} f_l ^{\rho \mu} -2
f_i ^{\mu \nu} f_j ^{\nu \sigma} f_k ^{\rho \mu} f_l ^{\sigma \rho} 
\nonumber 
\eea
The kinematical combination $C_{{\cal S}}$ coincides with the unique
kinematical invariant of the NS 4-point function encountered at tree and
1-loop level, which is often expressed in terms of the rank 8 tensor $t$
(see \cite{gs82, grosswitten}),
\be
\label{tee}
C_{\cal S} = - 8 t _{\kappa _1 \lambda _1 \kappa _2 \lambda _2
\kappa _3 \lambda _3 \kappa _4 \lambda _4} f _1 ^{\kappa _1 \lambda _1}
f_2 ^{\kappa _2 \lambda _2} f _3 ^{\kappa _3 \lambda _3} f _4 ^{\kappa _4
\lambda _4}
\ee
Finally, the forms ${\cal S}$ and ${\cal T}$ are given by
\bea
{\cal S} (1234) &=& 
- {1 \over 192 \pi ^6 \Psi _{10} } \ 
\omega _I (z_1) \omega _J (z_2) \omega _K (z_3) \omega _L(z_4) 
\sum _\delta \Xi _6 [\delta ] \tet [\delta ]^3 \p_I \p_J \p_K \p_L \tet
[\delta](\Omega)
\nonumber \\
{\cal T} (ij|kl) &=&
- {1 \over 8 \pi ^2} \ \omega _{[1} (z_1) \omega _{2]} (z_2) \omega _{[1}
(z_3) \omega _{2]} (z_4)
\eea
The $\delta$-sum for the ${\cal T}$-term was carried out explicitly, and
no $\Psi _{10}$ appears in its contribution. ${\cal S}$ and $C_{{\cal S}}$
are totally symmetric, while ${\cal T}$ and $C_{{\cal T}}$ are odd under
the interchange of $i\leftrightarrow j$ or $k\leftrightarrow l$. As a
result, the ${\cal T}$-term is novel at 2 loops and could not exist at 1
loop.

\medskip

The disconnected part of the $4$-point function for massless bosons,
calculated above, is finite.\footnote{This is the case at least when the
Mandelstam variables $k_i\cdot k_j$ are purely imaginary. As is now well
known \cite{dp94}, finiteness for general $k_i\cdot k_j$
cannot be read off directly, but has to be established by analytic
continuation.} Recall that the modular form $\Psi_{10}(\Omega)$ vanishes
of second order along the divisor of separating nodes. This corresponds to
the propagation of a tachyon, and was responsible
for divergence in the bosonic string \cite{div}.
Here however, the vector
\be
\sum_{\delta}
\Xi_6[\delta](\Omega)\tet ^3[\delta](\Omega)
\p_I \p_J \p_K \p_L \tet [\delta](\Omega)
\ee
also vanishes of second order along the divisor of
separating nodes, rendering the superstring amplitude
finite. The ${\cal A}$-term is manifestly finite.

\medskip

In the low energy limit, the exponential factor of the scalar Green's
function in (\ref{fourpoint}) tends to 1. It is instructive to identify
the kinematical factors that emerge from the integration over the 4
vertex insertion points $z_i$ of $| \F |^2$ in (\ref{fourpoint}) in the
Type II superstrings (analogous expressions may be derived for the
heterotic strings). The first contribution is from the product
$C_{\cal S} \bar C _{\cal S}$, and yields the well-known $t t R^4$ term of
four Riemann tensors contracted with two copies of the rank 8 tensor $t$
of (\ref{tee}) as obtained in \cite{grosswitten}. As argued in the
preceding paragraph, this contribution is given by a convergent integral.
The second contribution is from the products $C_{\cal S} \bar C_{\cal
T}$; it vanishes in view of the complete symmetry in the points $z_i$ in
${\cal S}$ and the antisymmetry in two pairs of points in ${\cal T}$. The
third contribution is from the product $C_{\cal T} \bar C_{\cal T}$ for
which the $z_i$ integrals may be carried out using the Riemann
bilinear relations. The resulting kinematical factors is again a
quadrilinear in the Riemann tensor and is proportional to
\bea
C_{\cal T} \bar C_{\cal T} & \longrightarrow &
+  \bigl (
R_{\alpha \beta \mu \nu} R^{\alpha \beta \mu \nu} \bigr )^2 
-  
R_{\alpha \beta \mu \nu} R^{\gamma \delta \mu \nu}
R^{\alpha \beta \rho \sigma } R_{\gamma \delta \rho \sigma}
\nonumber \\ &&
+ 4 
R_{\alpha \beta \mu \nu} R^{\gamma \delta \mu \nu} 
R^{\beta}{}_{ \gamma \rho \sigma} R_{\delta}{}^{ \alpha \rho \sigma}
- 4 
R^{\alpha \beta \mu \nu} R_{\delta \alpha \mu \nu} 
R_{\beta \gamma \rho \sigma} R^{\gamma \delta \rho \sigma}
\nonumber \\ &&
+ 4 
R^{\alpha \beta \mu \nu} R_{\beta \gamma \nu \rho}
R^{\gamma \delta \rho \sigma} R_{\delta \alpha \sigma \mu}
- 4 
R^{\alpha \beta \mu \nu} R_{\beta \gamma \nu \rho}
R_{\delta \alpha}{}^{ \rho \sigma} R^{\gamma \delta}{}_{ \sigma \mu}
\eea
While it is possible that this term, which arose from the
{\sl disconnected contributions} in (\ref{disconnected}), will be
cancelled by similar contributions arising from the {\sl connected
contributions} in (\ref{connected}), the above contribution to the low
energy effective action has at least one remarkable property~: the
integral over moduli space becomes simply the volume of moduli space with
respect to the $Sp(4,{\bf Z})$ invariant volume form $\prod _{I\leq J}
|d\Omega _{IJ} |^2 (\det \Im \Omega )^{-3}$. We note that the problem of
loop corrections in Type II superstrings and their contribution to low
energy effective actions has witnessed a resurgence of interest recently
(see for example
\cite{greenvanhove}).

\section{Compactification with worldsheet supersymmetry}
\setcounter{equation}{0} 

Our methods extend easily to
compactifications of some of the space-time directions
to a manifold $C$, under the
following basic assumptions

\begin{itemize}

\item The compactification only modifies the matter conformal field
theory, leaving the superghost part unchanged;

\item The compactification respects $\N=1$ local worldsheet supersymmetry,
so that the super-Virasoro algebra with matter central charge $c=15$ is
preserved.

\end{itemize}

Under these conditions, the superstring measure is independent of any
choices of gauge slice,  and a simple prescription for its
calculation can be given in terms of the OPE of two supercurrents.
Denote by $C$ resp. $M$ the fact that the corresponding object is
considered on the space-time manifold $C$ resp. Minkowski space $M$. In
particular, the chiral partition functions for the matter parts of $C$
and $M$ will be denoted by $Z_C$ and $Z_M$ respectively.  Then we have
the following result,
\bea
\label{comp}
{\cal A}_C[\delta] &= & {\cal A}_M
[\delta] {Z_C \over Z_M} \biggl \{
1 - {1 \over 8 \pi ^2} \int \! d^2 \! z \int \! d^2 \! w
\chiz \chiw [\< S_C (z) S_C (w) \> _C - \< S_M (z) S_M (w) \>_M ]
\nonumber \\
&& \hskip 1in 
+{1 \over 2 \pi} \int \! d^2 \! z \hat \mu _{\bar z} {}^z [\<T_C(z) \> _C
- \<T_M (z)\>_M ] \biggr \}
\eea
Here, $\hat \mu_{\bar z}^z$ is a Beltrami differential
shifting the complex structure from $\Omega_{IJ}$ to
$\hat\Omega_{IJ}$. The terms $S_C(z)$, $S_M(z)$, $T_C(z)$, $T_M(z)$
are the supercurrents and stress tensors. 
The expressions $\<S_C(z)S_C(w)\>$ and $\<T(z)\>$ are
the chiral correlation functions of $S(z)$ and $T(z)$,
or more precisely, the superconformal blocks of the
corresponding correlation functions.
They are related by the OPE,
\bea
S_C(z) S_C(w) & = & {1 \over 4} { T_C(z) + T_C(w) \over z-w} + (z-w) \O_C
(w)  + \O (z-w)^2
\nonumber \\
S_M(z) S_M(w) & = & {1 \over 4} { T_M(z) + T_M(w) \over z-w} + (z-w) \O_M
(w)  + \O (z-w)^2
\eea
for some non-universal operators $\O_C$ and $\O_M$.
The expression ${\cal A} _C [\delta]$
is independent of $\hat \mu_{\bar z}{}^z$ within its conformal
class since $\<T_C(z)\>_C-\<T_M(z)\>_M$ is a holomorphic 2-form.
The argument for the supercurrents is similar.
Indeed, since the ghost parts of $S_C(z)$ and $S_M(z)$ coincide,
all the singularities in $z$ and $w$ with the insertion points
$p_a$ and $q_{\alpha}$ are the same, and cancel between the $C$
and the $M$ contributions. Thus the only possible singularity in
the $SS$ correlator is when $z\to w$, and these contribute to
the supersymmetry variation $\delta_{\xi}\chiz=-2\p_{\bar z}\xi^+$.
Just as in the flat Minkowski case, this singularity is
cancelled precisely by the variation $\delta_{\xi} \hat \mu _{\bar z} {}^z
=\xi^+\chiz$, in view of the OPE. Thus $\<S_C(z)S_C(w)\>_C
-\<S_M(z)S_M(w)\>_M$ is singularity free, and ${\cal A}_C[\delta]$
is slice independent, just as ${\cal A}[\delta]$ was.

\medskip

It is then straightforward to evaluate ${\cal A} _C [\delta]$, for
example, by collapsing all points $p_a$ to a single point, or by going to
the gauge (\ref{split}). In particular, in the gauge (\ref{split}), we
find
\be
\label{orb}
{\cal A} _C [\delta]
=
{Z_C\over Z_M}
\bigg\{
{\cal Z}
+
{\zeta^1\zeta^2\over 16\pi^6}.
{\Xi_6[\delta]\tet[\delta](0)\over\Psi_{10}}
-
{\zeta^1\zeta^2\over 4\pi^2} 
{\cal Z}\<S_C(q_1)S_C(q_2)\>\bigg\}
\ee
where ${\cal Z}$ is the basic Minkowski space-time matter - ghost
correlator 
\be
{\cal Z}=\< \prod _a b(p_a) \prod _\alpha \delta (\beta
(q_\alpha)) \> /\det \bigl (\omega _I \omega _J (p_a) \bigr )
 \cdot \det \< \chi _\alpha | \psi ^* _\beta\>
\ee
The model of \cite{ks} is given by compactification
on an orbifold. The above formulas provide an explicit and consistent
framework for the calculation of the cosmological constant in this model
and others.

\bigskip

\noindent
{\large \bf  Acknowledgements}

\medskip

We are happy to acknowledge conversations with Zvi Bern, Per Kraus, Boris
Pioline and Pierre Vanhove. We are especially grateful to Costas Bachas
and Edward Witten for their continued interest in this work and for their
encouragement throughout. We wish to thank the Aspen Center for Physics,
the Ecole Normale Sup\' erieure, the Institut Henri Poincar\' e, and the
National Center for Theoretical Science in Hsin-Chu, Taiwan for their
hospitality while part of this work was being carried out.

\end{document}